\documentclass[12pt,twoside]{article}
\usepackage{a4wide,latexsym,graphicx,epsfig,psfrag}
\usepackage{slashed}
\usepackage{graphics}
\usepackage{subfig}
\usepackage{epsfig}
\usepackage{amsmath}
\usepackage{amssymb}
\usepackage{array}
\usepackage{wrapfig}
\usepackage{amsmath,amssymb,hyperref,xspace}
\usepackage{mciteplus,color,mathtools,graphicx,mathabx}
\usepackage{bm,bbm,slashed}
\usepackage{multirow}
\usepackage[english]{babel}
\usepackage{mathrsfs}
\usepackage{enumerate}
\usepackage{xspace}
\usepackage{slashed}
\usepackage{wrapfig}
\usepackage[export]{adjustbox}
\usepackage[utf8]{inputenc}
\usepackage{subfig}
\usepackage[toc,page]{appendix}
\usepackage{array,multirow}
\usepackage{simplewick}
\usepackage[T1]{fontenc} 
\usepackage{longtable}
\usepackage{appendix}
\graphicspath{ {./images/} }
\usepackage{pdflscape}
\usepackage{booktabs}
\usepackage[sort&compress,numbers]{natbib}
\usepackage{longtable} 

\usepackage{siunitx}
\usepackage[dvipsnames]{xcolor}
\allowdisplaybreaks
{\catcode `\@=11 \global\let\AddToReset=\@addtoreset}
\AddToReset{equation}{section}

\def\greaterthansquiggle{\raise.3ex\hbox{$>$\kern-.75em\lower1ex\hbox{$\sim$}}}
\def\lessthansquiggle{\raise.3ex\hbox{$<$\kern-.75em\lower1ex\hbox{$\sim$}}}
\newcommand{\beq}{\begin{equation}}
\newcommand{\eeq}{\end{equation}}
\newcommand{\beqa}{\begin{eqnarray}}
\newcommand{\eeqa}{\end{eqnarray}}
\newcommand{\beqan}{\begin{eqnarray*}}
\newcommand{\eeqan}{\end{eqnarray*}}
\newcommand{\ba}{\begin{array}}
\newcommand{\ea}{\end{array}}

\newcommand{\nn}{\nonumber \\}
\newcommand{\bea}{\begin{eqnarray}}
\newcommand{\eea}{\end{eqnarray}}

\def\nz{\ifmmode {I\hskip -3pt N} \else {\hbox {$I\hskip -3pt N$}}\fi}
\def\zz{\ifmmode {Z\hskip -4.8pt Z} \else
       {\hbox {$Z\hskip -4.8pt Z$}}\fi}
\def\qz{\ifmmode {Q\hskip -5.0pt\vrule height6.0pt depth 0pt
       \hskip 6pt} \else {\hbox
       {$Q\hskip -5.0pt\vrule height6.0pt depth 0pt\hskip 6pt$}}\fi}
\def\rz{\ifmmode {I\hskip -3pt R} \else {\hbox {$I\hskip -3pt R$}}\fi}
\def\cz{\ifmmode {C\hskip -4.8pt\vrule height5.8pt\hskip 6.3pt} \else
       {\hbox {$C\hskip -4.8pt\vrule height5.8pt\hskip 6.3pt$}}\fi}

\def\au{{\setbox0=\hbox{\lower1.36775ex%
\hbox{''}\kern-.05em}\dp0=.36775ex\hskip0pt\box0}}
\def\ao{{}\kern-.10em\hbox{``}}

\normalsize
\begin{document}
\begin{titlepage}
\vspace{2.5cm}
\begin{flushright}
 INT-PUB-24-004
\\
\end{flushright}

\vspace{2.0cm}

\begin{center}
{\LARGE \bf 

{One-loop analysis of $\beta$ decays in SMEFT}
}
\\[40pt]

{\large \bf  Maria Dawid, Vincenzo Cirigliano, Wouter Dekens} 

\vspace{0.25cm}

\vspace{0.25cm}
{\large 
{\it 
Institute for Nuclear Theory,\\ University of Washington, Seattle WA 98195-1550, USA}}

\vspace{1cm}

{\bf Abstract} \\
\end{center}
\noindent

We perform a loop-level analysis of  charged-current  (CC)
processes involving light leptons and quarks 
within the Standard Model Effective Field Theory (SMEFT). 
This work is motivated by the high precision reached in 
experiment and Standard Model calculations for CC decays of mesons, 
neutron, and nuclei, and by a lingering tension in the Cabibbo universality test. 
We identify the SMEFT operators that induce the largest loop-level contributions to  CC processes. 
These include four-quark and four-fermion semileptonic operators involving two third-generation quarks. 
We discuss the available constraints on the relevant effective couplings and along 
the way we derive new loop-level bounds 
from $K \to \pi \nu \bar \nu$ on four-quark operators involving two top quarks. 
We find that low-energy CC processes are quite competitive
with other probes, 
set constraints that do not depend on flavor-symmetry assumptions, 
and probe operators involving  third-generation quarks up to effective scales of $\Lambda\simeq 8$ TeV.
Finally, we briefly discuss single-field ultraviolet completions that could induce the relevant operators.

\vfill

\end{titlepage}

\section{Introduction}
The Standard Model (SM) is a successful
theory describing the interactions of all known elementary particles, with its predictions 
validated by experimental 
data up to the TeV energy scale. Yet, the SM encounters both theoretical and experimental issues that it is unable to explain. 
This suggests that the SM is incomplete and, if the new physics resides at a high mass scale, should be seen as
the leading term of 
an effective field theory.
The resulting theory is the Standard Model Effective Field Theory (SMEFT) \cite{Buchmuller:1985jz,Grzadkowski:2010es}, which extends the SM with higher-dimensional operators that are suppressed by inverse powers of a new physics scale ($\Lambda$). 
This theoretical framework allows for a systematic exploration of potential deviations from the SM predictions in terms of new physics of ultraviolet origin. 

The SMEFT framework is widely used at the tree level to analyze data on precision tests of the SM or searches for rare/forbidden processes.
For certain observables, such as permanent electric dipole moments~\cite{McKeen:2012av,Brod:2013cka,Chien:2015xha,Cirigliano:2016nyn,Brod:2023wsh}, it has been common to employ SMEFT at the loop level
with the appropriate Renormalization Group Equations (RGEs),  in order to constrain operators involving heavy SM fields. 
The calculation of the full set of one-loop anomalous dimensions in SMEFT~\cite{Jenkins:2013zja,Jenkins:2013wua,Alonso:2013hga} 
has enabled the study of a broader set of observables at the loop level, including precision measurements (see, for example,~\cite{Aebischer:2021uvt,Greljo:2023bdy}). 
This more general type of analysis reveals the full constraining power of low-energy precision measurements, 
including sensitivity to operators involving heavy particles, such as the top quark and the Higgs boson.  
In this paper, we focus on a one-loop SMEFT analysis of semileptonic charged-current 
processes involving light quarks. 
This study is motivated by: 

(i) The high precision reached by experiment and 
SM theory predictions for $\beta$ decays of mesons, hadrons, and nuclei. 
In a tree-level analysis, the current per-mille level precision
corresponds to
probing effective scales up to $\Lambda \sim 20$~TeV.

(ii) The emergence of tensions with the
Cabibbo universality test, related to the unitarity of the Cabibbo-Kobayashi-Maskawa (CKM) matrix, 
which requires  $\Delta_{\rm CKM} \equiv |V_{ud}|^2+|V_{us}|^2+|V_{ub}|^2-1=0$. 
The current hints of $\Delta_{\rm CKM} \neq 0$ at $\sim 3\sigma$  
have generated scrutiny within the SM~\cite{Feng:2020zdc,Seng:2020wjq,Ma:2021azh,Yoo:2023gln,Cirigliano:2023fnz,Seng:2022cnq,Seng:2023cvt,Cirigliano:2022yyo,1100705,Seng:2022epj,Seng:2022inj,Ma:2023kfr,Seng:2023cgl}
as well as beyond the SM (BSM) studies~\cite{Belfatto:2019swo,Grossman:2019bzp,Crivellin:2020lzu,Kirk:2020wdk,Crivellin:2020ebi,Alok:2021ydy,Crivellin:2021bkd,
Crivellin:2022rhw,Belfatto:2021jhf,Belfatto:2023tbv,Gonzalez-Alonso:2016etj,Falkowski:2017pss,Cirigliano:2021yto,Cirigliano:2023nol}, 
in the context of explicit models of new physics,  in the EFT setting below the weak scale, and in the SMEFT at tree level.

In this work, we will identify the SMEFT operators that induce the largest loop-level contributions to 
semileptonic charged-current (CC) processes. 
We will subsequently use low-energy data to indirectly constrain their size and compare the results to other available
limits, from both low-energy and collider physics. 
As we discuss below,  $\beta$ decays are quite competitive 
and set constraints that are independent of any assumptions 
about flavor symmetries.

The paper is organized as follows. We discuss the SMEFT formalism and the operators relevant to low-energy CC processes in Section \ref{sec:formalism}. Section \ref{sec:results} discusses the size of the RGE contributions, the resulting constraints from low-energy CC processes, as well as the limits from other probes. We investigate the possible UV origins of the operators most severely constrained in Section \ref{sec:models}, before concluding in Section \ref{sec:conclusions}. Detailed results of the RGE solutions are relegated to the Appendix \ref{ap:results}.

\section{Formalism}\label{sec:formalism}
The Standard Model EFT Lagrangian consists of the $SU(3)_C \times SU(2)_L \times U(1)_Y$-invariant operators incorporating the Standard Model fields. It contains the Standard Model Lagrangian $\mathcal{L}_{\rm SM}$ along with higher-dimensional operators that capture the effects of heavy degrees of freedom with masses greater than $\Lambda$. Focusing on dimension-6 operators, we have
\begin{equation}
\mathcal{L}_{\rm SMEFT} = \mathcal{L}_{\rm SM} + \frac{1}{v^2} \sum_k C_k^{(6)} Q_k^{(6)} \ .
\end{equation} 
Here $Q_{k}^{(i)}$ denotes all possible operators of canonical dimension $i$ and $C_k^{(i)}$ are the corresponding Wilson coefficients. For convenience, we have defined the coefficients to be dimensionless, so that they scale as $C_k^{(i)}\sim v^2/\Lambda^2$, where $v\simeq 246$ GeV  is the Higgs vacuum expectation value.
In this work, we exclusively focus on dimension-6 operators, as higher order operators are further suppressed by higher powers of $1 / \Lambda$ making their contributions less significant relative to those of dimension six. Our approach includes a general flavor structure and employs the so-called Warsaw basis~\cite{Grzadkowski:2010es} which contains 2499 independent operators~\cite{Alonso:2013hga}.

The SMEFT Lagrangian is written in a weak eigenstate basis, where the fermion mass matrices are not diagonal. 
We work in the weak basis in which the mass matrices take the form~\cite{Jenkins:2013zja} 
~\footnote{We neglect the effects induced by neutrino masses. These are induced at dimension-5 in SMEFT \cite{Weinberg:1979sa}, but are negligible for our purposes.} 
\begin{align}\label{eq:basis}
M_u = \text{diag}(m_u, m_c, m_t) \ , ~
M_d = \text{diag}(m_d, m_s, m_b) \cdot V^{\dagger} \ , ~
M_e = \text{diag}(m_e, m_{\mu}, m_{\tau}) \ ,
\end{align}
where $V$ is the unitary CKM matrix.
In this basis the  left-handed down-type quarks flavor fields 
are given by $V d_L$ in terms of the  CKM matrix $V$ and 
the mass eigenfields $d_L$.
Hence, 
the fermion fields 
can be written in terms of the mass eigenstate fields as follows
\begin{equation*}
    q_i = \begin{pmatrix} u_{L}\\ V d_{L}\\\end{pmatrix}_i, ~~~~~~~~ l_i = \begin{pmatrix} \nu_{L}\\ e_{L}\\\end{pmatrix}_i, ~~~~~~~~ u_i=u_{R_i}, ~~~~~~~~ d_i=d_{R_i}, ~~~~~~~~  e_i=e_{R_i} \ .
\end{equation*}

The complete set of dimension-six SMEFT operators 
contributing to $\beta$ decay (and more generally to semileptonic CC processes)  at  tree level is 
given by~\cite{Cirigliano:2012ab}
\begin{subequations}
\label{eq:beta-operators}
\begin{eqnarray}
&& 
\text{\underline {Four-fermion operators:}}~~~~~~~~~~~~~~~~~~~~~~~~~~~~
\nonumber\text{\underline {Vertex corrections:}}\\
&& \label{eq:1}
Q_{\substack{lq\\prst}}^{(3)} = (\bar l_p \gamma_\mu \tau^I l_r)(\bar q_s \gamma^\mu \tau^I q_t)~~~~~~~~~~~~~~~~~~~~~~~~
Q_{\substack{Hl\\pr}}^{(3)} = (H^\dag i\overleftrightarrow{D}^I_\mu H)(\bar l_p \tau^I \gamma^\mu l_r)~~~~~~~~
\\
&&\label{eq:2}
Q_{\substack{ledq\\prst}} = (\bar l_p^j e_r)(\bar d_s q_{tj})~~~~~~~~~~~~~~~~~~~~~~~~~~~~~~~~~
Q_{\substack{Hq\\pr}}^{(3)} = (H^\dag i\overleftrightarrow{D}_\mu H)(\bar q_p \gamma^\mu q_r)~~~~~~~~~~
\\
&&\label{eq:3}
Q_{\substack{lequ\\prst}}^{(1)} = (\bar l_p^j e_r) \epsilon_{jk} (\bar q_s^k u_t) ~~~~~~~~~~~~~~~~~~~~~~~~~~~~~~
Q_{\substack{Hud\\pr}} = i(\widetilde H ^\dag D_\mu H)(\bar u_p \gamma^\mu d_r)~~~~~
\\
&&\label{eq:4}
Q_{\substack{lequ\\prst}}^{(3)} = (\bar l_p^j \sigma_{\mu\nu} e_r) \epsilon_{jk} (\bar q_s^k \sigma^{\mu\nu} u_t)~. ~~~~~~~~~~~
~~~~~~
\end{eqnarray}
\end{subequations}
Here $D_{\mu}=   I \, \partial_\mu   \, + \, i g_3 T^A A_{\mu}^A \, + \, i g_2 \frac{\tau^I}{2} W_\mu^I   \, +\, i g_1  Y B_\mu$ is the gauge covariant derivative,  $T^A$ are the $SU(3)$ generators, $\tau^{I}/2$ are the $SU(2)$ generators, $Y$ is the $U(1)$ hypercharge, while $B_\mu$, $A_\mu^A$, and $W^I_\mu$ are the gauge fields and $g_1$, $g_2$, $g_3$ the corresponding gauge couplings.
The Higgs field in the unitary gauge is given by $H=\sqrt{1/2} (0,\, h+v)^T$, with $\widetilde H = i\tau_2 H^*$. Finally, the combinations of covariant derivatives are given by $\overleftrightarrow{D}_\mu = \overrightarrow{D}_\mu -\overleftarrow{D}_\mu$ and $\overleftrightarrow{D}^I_\mu =\tau^I \overrightarrow{D}_\mu -\overleftarrow{D}_\mu \tau^I$, while the subscripts $p,\ r,\ s,\ t$ represent flavor indices.

\subsection{Low-scale Effective Lagrangian and tree level bounds}

In order to describe the new physics contributions to semileptonic CC decays 
of mesons, neutron, and nuclei, we need to evolve the effective Lagrangian involving the operators 
in Eq.~(\ref{eq:1}) down to a scale of $O({\rm GeV})$. To do so, 
at the weak scale $\mu \sim m_W$, we match the SMEFT to
the so-called low-energy Effective Field Theory (LEFT) \cite{Jenkins:2017jig},
which is invariant under $SU(3)_c \times U(1)_{em}$ and contains the Standard Model 
fields except for the heavy $W$, $Z$, and $h$ bosons as well as the top quark. 
The LEFT Lagrangian for the semileptonic $d_j \to u$ transitions, with $d_1= d$ and $d_2 = s$, 
is given by
\cite{Cirigliano:2009wk,Cirigliano:2012ab,Gonzalez-Alonso:2016etj,Falkowski:2020pma}~\footnote{We write here 
only the operators involving electrons and their neutrinos. The inclusion of the second family of leptons
is straightforward. Our analysis below assumes lepton family universality.} 
\begin{eqnarray}
\label{LEFT_beta_Lagrangian}
\mathcal{L}_{\rm LEFT} =&-&\frac{G_F V_{ud_j}}{\sqrt{2}} \Big[ \left(1+\epsilon_L^{\beta d_j} - \epsilon_L^\mu \right) \ \bar{e}\gamma_{\mu}(1-\gamma_5)\nu_e \cdot \bar{u}\gamma^{\mu}(1-\gamma_5)d_j 
\nonumber \\
&+&\epsilon_R^{d_j} \  \bar{e}\gamma_{\mu}(1-\gamma_5)\nu_e \cdot \bar{u}\gamma^{\mu}(1+\gamma_5)d_j   \nonumber \\ 
&+&\epsilon_S^{d_j} \ \bar{e}(1-\gamma_5)\nu_e \cdot\bar{u}d_j - \epsilon_P^{d_j}\bar{e} (1-\gamma_5)\nu_e \cdot \bar{u}\gamma_5d_j   \nn 
&+& \epsilon_T^{d_j} \  \bar{e}\sigma_{\mu\nu}(1-\gamma_5)\nu_e \cdot \bar{u} \sigma^{\mu\nu}(1-\gamma_5) d_j  \Big] \ + 
{\rm h.c.} \ ,
\end{eqnarray}
where $G_F$ is the Fermi constant and the BSM contributions are encoded in the dimensionless couplings $\epsilon_i$. 
$\epsilon_L^{\mu}$ arises from the SMEFT correction to $G_F$ as extracted from muon decay and is given by
\begin{eqnarray}
  \epsilon_L^{\mu} = - C_{_{\substack{l l \\ 1221}}} \ + \left( C^{(3)}_{\substack{H l \\ 11}}+C^{(3)}_{\substack{H l \\ 22}} \right) \ ,
\end{eqnarray}
in terms of leptonic vertex corrections introduced  above and the coefficient associated with 
the purely leptonic operator (note that $C_{\substack{ll \\ prst}}=C_{\substack{ll \\ stpr}}$)
\begin{equation*}
    Q_{\substack{l l \\ prst}} = (\bar l_p \gamma_\mu l_r)(\bar l_s \gamma^\mu l_t) \ ~.
\end{equation*}

In order to connect the Wilson coefficients of SMEFT and LEFT, 
we require the equality of the $d_j \to u e^- \bar{\nu}_e$ amplitudes
at the energy scale $\mu \sim m_W$. 
The resulting matching conditions are~\cite{Cirigliano:2012ab}
\begin{subequations}
\label{eq:tree_level_matching}
    \begin{eqnarray}
& & \epsilon_L^{\beta d_j} = C^{(3)}_{\substack{H l\\ 11 }} +  \frac{[ C^{(3)}_{Hq}V]_{1j}}{V_{ud_j}} 
 -\frac{[C^{(3)}_{lq}V]_{1j} }{ V_{ud_j}} \ , \\
& &\epsilon_R^{d_j} = \frac{1}{2} \frac{[C_{\substack{Hud}}]_{1j}}{V_{ud_j}} , \\
& &\epsilon_S^{d_j} = - \frac{[C^{\dagger}_{\substack{ledq}} + C^{(1)\dagger  }_{\substack{lequ}}V]_{1j}}{2 \ 
V_{ud_j}} \ , \\
& &\epsilon_P^{d_j} = \frac{[C^{\dagger}_{\substack{ledq}} -  C^{(1)\dagger}_{\substack{lequ}}V]_{1j}}{2 \ 
V_{ud_j}} \ , \\
& &\epsilon_T^{d_j} = -\frac{[ C^{(3)\dagger}_{\substack{lequ}}V]_{1j}}{2 \ V_{ud_j}}\ . 
\end{eqnarray}
\end{subequations}
Semileptonic decays depend on the individual effective couplings $\epsilon_{R,S,P,T}^{d_j}$ and on the linear combination $\epsilon_L^{(d_j)} \equiv \epsilon_L^{\beta d_j} - \epsilon_L^{\mu}$, 
which can be written as follows in terms of the SMEFT Wilson coefficients: 
\begin{equation}
   \epsilon_L^{(d_i)} = \frac{[ C^{(3)}_{Hq}V]_{1j}}{V_{ud_j}} - \frac{[C^{(3)}_{lq}V]_{1j} }{V_{ud_j}} 
   + C_{_{\substack{l l \\ 1221}}}  - C^{(3)}_{\substack{H l \\ 22}} \ .
\label{eq:epsilonLdef}
\end{equation}
From the above equation, it follows that contributions of  $C^{(3)}_{\substack{H l \\ 11}}$ from muon and $\beta$ decay  cancel each other.

The rotation to the mass basis generates Wilson coefficients involving all family indices,  e.g., $ [C_{H q}^{(3)} \cdot V]_{11} = C_{\substack{H q \\ 11}}^{(3)}V_{11} + C_{\substack{H q \\ 12}}^{(3)}V_{21} + C_{\substack{H q \\ 13}}^{(3)}V_{31}$. In our analysis, we omit terms involving off-diagonal CKM matrix elements, as their contributions are 
suppressed by one or more powers of the Wolfenstein parameter $\lambda \equiv |V_{us}|  \sim 0.22$.
Hence, 
we 
identify the set of
of Wilson coefficients associated with the operators that contribute to semileptonic 
decays of mesons, baryons, and nuclei  at a tree-level ($j=1,2$):
\begin{equation}
\label{eq:Cbeta}
C_\beta = \{ C^{(3)}_{\substack{l q \\ 111j}} , C_{\substack{ledq \\ 11j1}},  C^{(1)}_{\substack{lequ \\ 11j1}},  C^{(3)}_{\substack{lequ \\ 11j1}},  C^{(3)}_{\substack{H l \\ 11}}, C^{(3)}_{\substack{H l \\ 22}},  C^{(3)}_{\substack{H q \\ 1j}},  C_{\substack{Hud\\ 1j}}, C_{\substack{l l \\ 1221}}, C_{\substack{l l \\ 2112}} \}. 
\end{equation}

Finally, to assess their impact on low-energy CC processes, the $\epsilon_i$ couplings need to be evolved from the weak scale down to hadronic scales, $\mu_0\simeq 2$ GeV. This leads to significant effects for the scalar and tensor operators, which have nonzero ${\cal O}(\alpha_s)$ anomalous dimensions. At one loop in QCD, we find $\epsilon_{S,P}(\mu_0)\simeq 1.64\, \epsilon_{S,P}(\mu_W)$ and $\epsilon_{T}(\mu_0)\simeq 0.85\, \epsilon_{T}(\mu_W)$, for $\mu_W=246$ GeV. The remaining couplings are subject to smaller, \%-level, effects due to QED loops, see Ref.\ \cite{Gonzalez-Alonso:2017iyc} for more details.

\subsection{Operator Mixing}
\begin{figure}[t!]%
    \centering
\includegraphics[height=5cm]{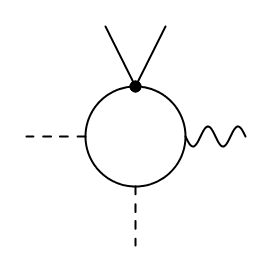}%
\hspace{0.8cm}
\includegraphics[height=5cm]{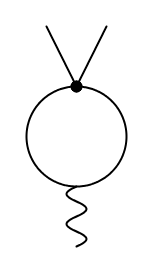} %
\hspace{0.8cm}
\includegraphics[height=5cm]{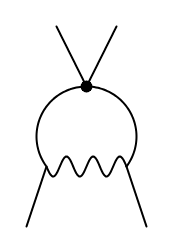} 
    \caption{Diagrams illustrating the loop contributions of $Q_x$ operators  to $Q_{\beta}$. 
    Black circles represent the insertion of $Q_x$ operators. 
    Solid lines denote fermions, dashed lines represent Higgs bosons, and wavy lines stand for gauge bosons.}%
    \label{fig:diagramy}%
\end{figure}

In the previous section, we expressed the effective couplings $\epsilon_i$ in terms of the SMEFT Wilson coefficients at the matching scale $\mu_W \sim M_W$ between SMEFT and LEFT. The next step is to evolve the SMEFT coefficients from the high energy scale $\Lambda$ to $\mu_W$ using the Renormalization Group Equations (RGEs). As the renormalization scale $\mu$ changes, the effective interactions mix, leading to the emergence of new operators in the beta decay analysis at the loop level. The RGEs for the Wilson coefficients of the relevant operators take the following form
\begin{equation}
\begin{bmatrix}
\dot{C_{\beta}}  \\
\dot{C_{x}}
\end{bmatrix} \  = \begin{bmatrix}
\gamma_{\beta \beta} & \gamma_{\beta x} \\
 \gamma_{x \beta } &  \gamma_{x x} 
\end{bmatrix}
\begin{bmatrix}
C_{\beta}  \\
C_{x} 
\end{bmatrix} \ ,
\end{equation}
where $\dot{C}_i = 16\pi^2 \mu \frac{d C_i}{d\mu}$ and $\gamma_{\beta \beta}$, $\gamma_{\beta x}$, $\gamma_{x x}$ are matrices of anomalous dimensions. 
$C_{\beta}$ is the array of Wilson coefficients 
associated with operators that contribute to semileptonic processes at the tree level, 
defined explicitly in Eq.~(\ref{eq:Cbeta}).
$C_x$ contains all the remaining Wilson coefficients that can mix onto the $C_\beta$'s,
i.e. for which $\gamma_{\beta x} \neq 0$.
Inspection of the one-loop RGEs~\cite{Jenkins:2013wua, Jenkins:2013zja, Alonso:2013hga} shows that
mixing occurs through the gauge and Yukawa interactions (we retain only the top- and bottom-quark Yukawa couplings). 
Numerically, the largest mixing occurs for coefficients  $C_x$
belonging to these classes: 
\begin{equation}
C_x \in \{ C^{(3)}_{ \substack {lq} },
C_{ \substack {ll} },
C^{(3)}_{\substack{H l}},
C^{(3)}_{\substack{H q}},
C^{(1)}_{ \substack {qq} },
C^{(3)}_{\substack{qq}},
C_{H \Box} \}~, 
\label{eq:Cx}
\end{equation}
with various combination of family indices, as discussed below. 
The most relevant one-loop diagram topologies  are given in Fig.~\ref{fig:diagramy}.

The coefficients $C_{\beta} (\mu_W)$ at a low scale close to the weak scale, 
expressed 
in terms of $C_x (\Lambda)$, 
take the form
\beq
C_{\beta i} (\mu_W) = \sum_k U_{ik}(\mu_W, \Lambda) C_{x k}(\Lambda) \ ,
\label{eq:CbetaRG}
\eeq
where the matrix $U_{}(\mu_W, \Lambda)$ is  obtained by solving the  RGEs. 
Consequently, 
by combining the mixing induced by RGE running
with the matching in Eq.\ \eqref{eq:tree_level_matching},
we can represent the  effective couplings $\epsilon_i$ 
in terms of the $C_x$ coefficients as
\beq
\epsilon_i(\mu_W) = \sum_j \kappa_{ij}(\mu_W, \Lambda) C_{x j}(\Lambda) \ . 
\label{eq:RGEsol}
\eeq
This formula, together with the experimental determination of the $\epsilon_i$, allows us to obtain constraints on $C_x (\Lambda)$. 
In this work, we take $\Lambda = 5$ TeV for the initial scale of the RG 
evolution~\footnote{Larger (smaller) values of $\Lambda$ will 
make the bounds we discuss below stronger (weaker). The dependence on $\Lambda$ is 
logarithmic, hence relatively weak.}. 
The results are presented and discussed in the following section.\\

\section{Results}\label{sec:results}

We find that the largest impact of operator running and mixing manifests  
in the low-energy coupling  $\epsilon_L (\mu_W)$. 
The mixing contributions to $\epsilon_{R,S,P,T}$ are less prominent 
and lead to weaker constraints on  the $C_x$.

\subsection{Right-handed  current, scalar, pseudoscalar, and tensor terms}

For this class of couplings, most of the non-zero mixings are small because of the appearance of the 
numerically suppressed $y_b$ Yukawa coupling. As an example, consider the evolution of the Wilson coefficient $C_{\substack{Hud}}$ associated with the right-handed quark current operator defined in Eq.~\eqref{eq:3}. The corresponding RGE term describing this evolution is 
\begin{align}
\dot C_{\substack{Hud  \\ 11}} & \supset 4(  C^{(1)}_{\substack{ud \\ 1331 }}+ \frac{4}{3}C^{(8)}_{\substack{ud \\ 1331 }})[Y_u Y_d^\dagger]_{33}
\end{align}
By solving the RGEs, we obtain a bound of $O(1)$ for the coefficients $ C^{(1),(8)}_{\substack{ud \\ 1331 }}$, implying a rather weak limit on the BSM scale, $\Lambda\gtrsim v$.
Analogous or even smaller results occur for the other $C_x$ coefficients that do not generate the left-handed currents. 

The results for $\epsilon^{(d)}_i$ at the scale $\mu_W = 246 $ GeV in terms of $C_{\beta}$ and $C_x$ SMEFT coefficients at $\Lambda = 5$  TeV are given by
\begin{eqnarray} \label{eq:epsilonR}
    \epsilon_R^{(d)}  &=& 0.950 \ C_{\substack {Hud\\11}} -  0.001 \  C^{(1)}_{\substack {ud\\1331}}\ -0.0011 \ C^{(8)}_{\substack {ud\\1331}}\ ~. 
\end{eqnarray}
\begin{eqnarray} \label{eq:epsilonP}
    \epsilon_P^{(d)}  &=& 0.603 \ C^{(1)}_{\substack {lequ\\1111}} - 0.600 \  C_{\substack {ledq\\1111}}\ + 0.143 \ C^{(3)}_{\substack {lequ\\1111}}\ ~. 
\end{eqnarray}
\begin{eqnarray} \label{eq:epsilonS}
    \epsilon_S^{(d)}  &=& - 0.603 \ C^{(1)}_{\substack {lequ\\1111}} - 0.600 \  C_{\substack {ledq\\1111}}\ + 0.143 \ C^{(3)}_{\substack {lequ\\1111}}\ ~.
\end{eqnarray}
\begin{eqnarray} \label{eq:epsilonT}
    \epsilon_T^{(d)}  &=& 0.003 \ C^{(1)}_{\substack {lequ\\1111}} - 0.465 \   C^{(3)}_{\substack {lequ\\1111}}\ ~.
\end{eqnarray}
\subsection{Operators involving left-handed currents}
The RGEs induce contributions from several Wilson coefficients to $\epsilon_L$. 
The coefficients of these operators are collected in $C_x$ in Eq.\ \eqref{eq:Cx}. Of the appearing operators, $Q_{\substack{ll}}$, $Q^{(3)}_{\substack{lq}}$, $Q^{(3)}_{\substack{Hl}}$, $Q^{(3)}_{\substack{Hq}}$ are found within the set of $Q_{\beta}$ operators in Eqs.\ \eqref{eq:1}-\eqref{eq:4}, but with a distinct flavor content from the operators that contribute at the tree level.
The other three operators are
\begin{subequations}
\begin{eqnarray}
\label{eq:Qqq1}&&Q^{(1)}_{ \substack {qq\\prst}} = (\bar q_p \gamma_\mu \tau q_r)(\bar q_s \gamma^\mu q_t),\\ \label{eq:Qqq3}
&&Q^{(3)}_{\substack{qq\\prst}} = (\bar q_p \gamma_\mu \tau^I q_r)(\bar q_s \gamma^\mu \tau^I q_t),\\
&&Q_{H \Box} = (H^{\dag} H)\Box(H^{\dag} H) \ .
\label{eq:Cx_form}
\end{eqnarray} 
\end{subequations}
In particular, the largest contributions to $\epsilon_i$, namely with $\kappa_{ij} \ge 10^{-2}$ (see \eqref{eq:RGEsol}) arise 
from 
$C^{(3)}_{\substack{qq \\ 1133}}$, $C^{(3)}_{\substack{qq \\ 1331}}$,$C^{(1)}_{\substack{qq \\ 1331}}$, $C_{\substack{ll \\ 1122}}$,  $C^{(3)}_{\substack{lq \\ 2233}}$, $C^{(3)}_{\substack{lq \\ 1111}}$, $C^{(3)}_{\substack{lq \\ 1122}}$,$C^{(3)}_{\substack{lq \\ 1133}}$, $C^{(3)}_{\substack{lq \\ 2211}}$, 
and $C^{(3)}_{\substack {Hq\\11}}$, $\ C^{(3)}_{\substack {Hq\\22}}$, $\ C^{(3)}_{\substack {Hq\\33}}$.

The numerical results for the RG evolution matrix  $U_{ik} (\mu_W, \Lambda)$ connecting 
$C_{xk} (\Lambda)$ to $C_{\beta i} (\mu_W)$  (see Eq. \eqref{eq:CbetaRG}) 
can be found in Table \ref{tab:U_results} in Appendix~\ref{ap:results}. 
Upon matching on the LEFT coefficients (see Eqs.(\eqref{eq:tree_level_matching}) we find that 
the $\epsilon^{(d,s)}_L$ at the scale $\mu_W = 246$~GeV  in terms of SMEFT coefficients at $\Lambda =5$~TeV 
are  given by
\begin{eqnarray} \label{eq:epsilonL}
    \epsilon_L^{(d)}  &=& 0.946 \ C^{(3)}_{\substack {Hq\\11}} -  1.052 \ C^{(3)}_{\substack {lq\\1111}} + 1.012 \ C_{_{\substack{l l \\ 1221}}} \ - 0.965 \ C^{(3)}_{\substack{H l \\ 22}} \\ 
    &+& \ 0.198 \ C^{(3)}_{\substack {qq\\1133}} - 0.026 \ C^{(3)}_{\substack {qq\\1331}} + 0.026 \ C^{(1)}_{\substack {qq\\1331}} - 0.0471 \ C_{\substack {ll\\1122}}  -  0.100 \ C^{(3)}_{\substack {lq\\2233}} ~, \nonumber
\end{eqnarray}
\begin{eqnarray} \label{eq:epsilonLs}
  \epsilon_L^{(s)}  &=& \frac{0.942}{\lambda} \ C^{(3)}_{\substack {Hq\\12}} -  \frac{1.052}{\lambda} \ C^{(3)}_{\substack {lq\\1112}}  + 1.010 \ C_{_{\substack{l l \\ 1221}}} \ - 0.960 \ C^{(3)}_{\substack{H l \\ 22}} \\ 
    &+& \frac{0.198}{\lambda} \ C^{(3)}_{\substack {qq\\1233}} - \frac{0.028}{\lambda} \ C^{(3)}_{\substack {qq\\1332}} + \frac{0.026}{\lambda} \ C^{(1)}_{\substack {qq\\1332}} - 0.099 \ C^{(3)}_{\substack {lq\\2233}} - 0.047  \ C_{\substack {ll\\1122}} \nonumber \\ 
    &+&  0.0144 \ (C^{(3)}_{\substack {Hq\\11}} + \ C^{(3)}_{\substack {Hq\\22}} + \ C^{(3)}_{\substack {Hq\\33}}) - 0.016 \ (C^{(3)}_{\substack {lq\\1111}} + C^{(3)}_{\substack {lq\\1122}} + C^{(3)}_{\substack {lq\\1133}}) ~. \nonumber 
\end{eqnarray}
Here $\lambda \equiv |V_{us}|  \sim 0.22$. 
To arrive at the expressions above, we used  the  symmetry under the exchange of
the fermion bilinears, 
$C^{(1),(3)}_{\substack{qq \\ prst}}=C^{(1),(3)}_{\substack{qq \\ stpr}}$, $C_{\substack{ll \\ prst}}=C_{\substack{ll \\ stpr}}$.
Note that $\epsilon_L^{(d)}$ and  $\epsilon_L^{(s)}$
have similar dependencies on  the $C_{x}$ Wilson coefficients in the first two lines of Eqs.\ \eqref{eq:epsilonL} and \eqref{eq:epsilonLs}, with appropriate replacements of 
the first and second family indices. The small \%-level differences   arise 
from two-loop level effects  captured by the RGEs in the leading-log approximation.

The expressions in Eqs.\ \eqref{eq:epsilonR}-\eqref{eq:epsilonT} as well as \eqref{eq:epsilonL} and \eqref{eq:epsilonLs} are in principle modified by one-loop corrections to the tree-level matching in Eq.\ \eqref{eq:tree_level_matching}. As we do not control all contributions at this order, i.e.\ one-loop terms without large logarithms $\sim \ln(\Lambda/m_W)$, they are beyond the precision of our leading-log analysis. Nevertheless, using the results of Ref.\ \cite{Dekens:2019ept}, we estimated the order of magnitude of these terms, finding that they do not lead to large contributions of new Wilson coefficients in the above equations. In addition, the corrections to the $C_x$ contributions found above tend to be small compared to our leading-log results. For example, the loop-level matching affects the contributions of  $[C_{qq}^{(3)}]_{1j33}$ in Eqs.\ \eqref{eq:epsilonL} and \eqref{eq:epsilonLs} by ${\cal O}(10\%)$.

To impose constraints on the difference $\epsilon_L^{(d_j)} = \epsilon_L^{\beta d_j} - \epsilon_L^{\mu}$, we employ the analysis of~\cite{Cirigliano:2023nol}. That reference utilized the first-row unitarity relation for the Cabibbo-Kobayashi-Maskawa (CKM) matrix by incorporating data from kaon, pion, and $\beta$ decay, following Refs.~\cite{Gonzalez-Alonso:2016etj,Falkowski:2020pma}. Apart from these low-energy CC processes, this analysis included electroweak precision measurements and collider observables, thereby allowing for a nearly model-independent analysis of SMEFT operators. However, as argued above, the effects of the $C_x$ coefficients mostly induce the left-handed interactions governed by $\epsilon_L^{(d_j)}$. Therefore, we perform a simpler analysis in which only these couplings are turned on.
The constraints on $\epsilon_{L}^{(s,d)}$ 
are dominated by the  CKM unitarity test, which implies 
\begin{equation} \label{eq:deltaCKM}
    \Delta_{\rm CKM} = 2 |V_{ud}|^2\epsilon_L^{(d)} + 2 |V_{us}|^2 \epsilon_L^{(s)}\,.
\end{equation}
From Eqs.~\eqref{eq:epsilonLs} and \eqref{eq:deltaCKM}
one sees that  $C_x$'s that contribute to  $\epsilon^{(s)}_L$  
appear in $\Delta_{\rm CKM}$ with at least one extra power of $\lambda$ compared to the 
$C_{xk}$ that contribute to $\epsilon^{(d)}_L$.
So  in what follows we focus on the strongest constraints on  the $C_{xk}$'s appearing in Eq.~\eqref{eq:epsilonL} provided by 
\begin{eqnarray}
    \epsilon_L^{(d)}  = (8.3 \pm 2.5)\times 10^{-4}\,.
\label{eq:epsLbound}\end{eqnarray}
The  constraints on the coefficients appearing in \eqref{eq:epsilonLs}  can be obtained via  rescaling by $\lambda$.

\begin{figure}[t!]
\centerline{\includegraphics[scale=.45]{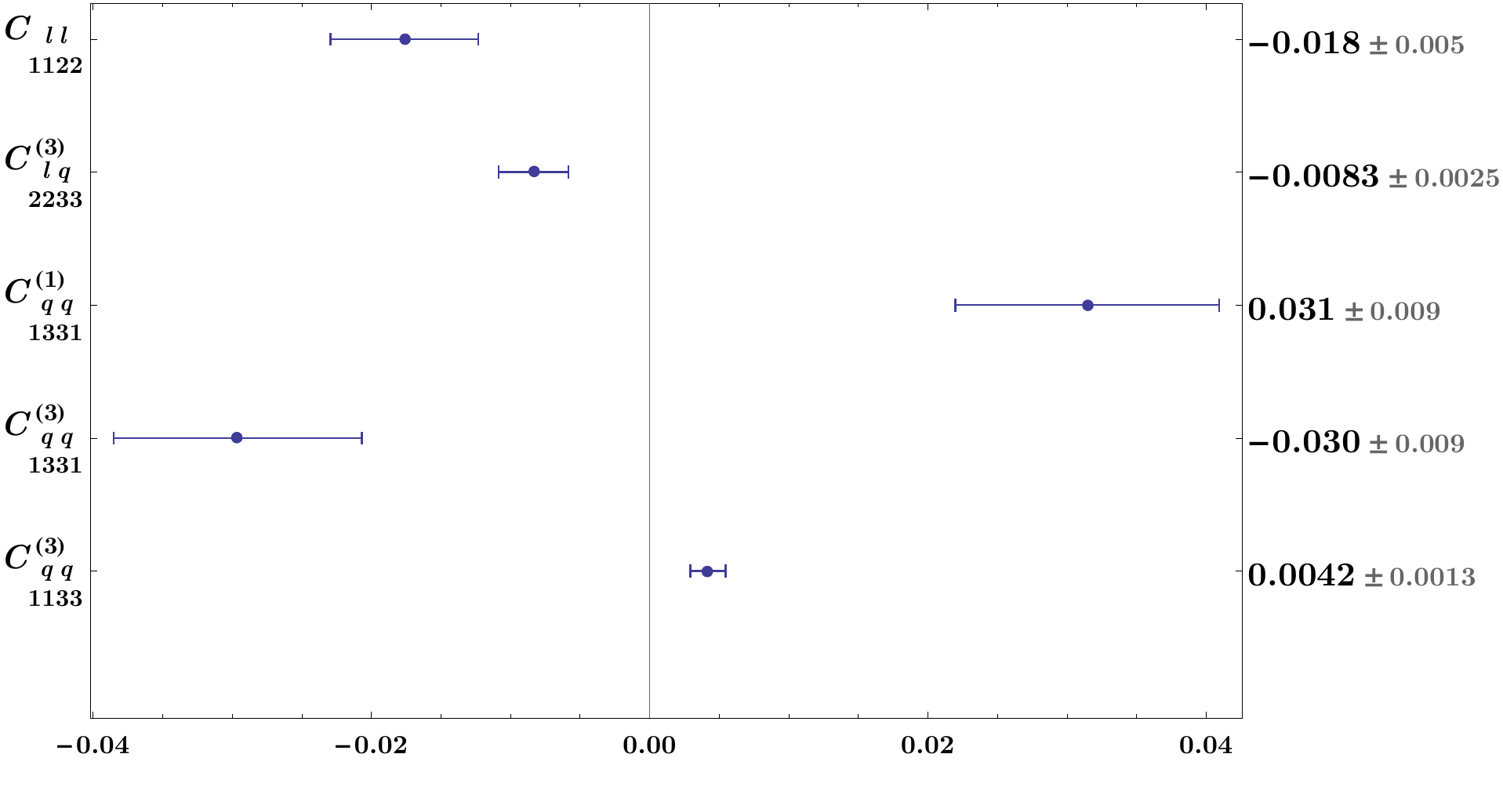}}
\caption{Summary of the strongest low-energy CC constraints on 
the coefficients $C_{xk} (\Lambda)$ ($68\%$ C.L.). 
We only show results for the most strongly constrained coefficients, 
with a threshold of  $\delta C_{xk} < 5 \times 10^{-2}$.}
\label{fig:wyniki}
\end{figure}

While Eq.~\eqref{eq:epsilonL} provides the general contributions, we explore below the simplified scenario in which only one coefficient is nonzero at the high energy scale $\Lambda$. This leads to conservative constraints on the $C_x$ operators, as long as there are no significant cancelations between different SMEFT operators. 
The most stringent constraints (corresponding $\delta C_x \leq 5 \times 10^{-2}$)
are graphically illustrated in Fig.~\ref {fig:wyniki} 
and reported in Table \ref{tab:results},   
along with a comparison to  limits obtained from other observables.

As can be seen from the Table, low-energy CC processes probe some of the Wilson coefficients at the permille level. 
For the purely leptonic operator, this leads to a sensitivity comparable to that of muon pair production. 
For semileptonic  operators involving the top quark, the constraints of low-energy CC processes 
are stronger than the ones from top production, but weaker than the ones 
from  rare $\Delta F=1$ meson decays and the $K_L-K_S$ mass difference. 
It should be noted, however,  that the constraints from the $\Delta F=1$ and $\Delta F=2$ processes depend on the flavor structure of the Wilson coefficients. In particular, defining the Wilson coefficients in a basis where $M_d$, instead of $M_u$, is diagonal would lead to far less stringent constraints from these probes, while those from low-energy CC measurements would be mostly unchanged.
We provide details on our analysis of $\Delta F=1$ and $\Delta F=2$ processes 
in Section~\ref{sect:fcnc}.

 \begin{table}[!t]
  \begin{adjustbox}{center,max width=\linewidth}
  \def\arraystretch{3}%
    \begin{tabular}{|c|c|c|c|}
\hline
      $C_{x} (\Lambda=5 \,{\rm TeV})$  &  Constraint from  & Strongest constraints& Process  \\
                                    &   $\beta$ decays &  from other processes &     \\
\hline 
      \hline
      $C^{(3)}_{\substack{q q \\ 1 1 3 3}}$  & $0.004 \pm 0.0013$  & $-0.0073 \pm 0.006^{(1)}$     & Top production~\cite{ Kassabov:2023hbm}     \\
        &&$\pm 0.00002^{(2)}$ & $K \to \pi \nu \bar \nu$ \\
      \hline
      $C^{(3)}_{\substack{l q \\ 2 2 3 3}}$  & $-0.008 \pm 0.0025$ & 
      $-0.00024 \pm 0.00021^{(2)}$
      & B decays ($R_K$)~\cite{Garosi:2023yxg}           \\
      \hline
      $C_{\substack{l l \\ 1 1 2 2}}$        & $-0.018 \pm 0.005$  & $- 0.0018 \pm 0.0029$   & Muon pair production~\cite{Falkowski:2015krw} \\
      \hline
      $C^{(1)}_{\substack{q q \\ 1 3 3 1}}$  & $-0.031 \pm 0.009$    &   
      $-0.035 \pm 0.027^{(1)}$
      & Top production~\cite{ Kassabov:2023hbm}        \\
      &&$\pm 0.0009^{(2)}$ & $\Delta m_K$~\cite{Aebischer:2020dsw}  \\
      \hline
     $C^{(3)}_{\substack{q q \\ 1 3 3 1}}$  & $-0.03 \pm 0.009$    & 
      $-0.042 \pm 0.024^{(1)}$
      & Top production~\cite{ Kassabov:2023hbm}     \\
         &&$\pm 0.00004^{(2)}$ & $K \to \pi \nu \bar \nu$\\
\hline 
    \end{tabular}
  \end{adjustbox}
  \caption{ 
Summary of the strongest  loop-level constraints from $\beta$ decays, 
defined by the threshold $\delta C_{xi} (\Lambda=5~\rm{TeV}) \leq 0.05$. 
We display the relevant SMEFT Wilson coefficients (first column), the 
$68\%$ C.L. constraint from $\beta$ decays (second column), and the strongest  
$68\%$ C.L.
constraints from other processes (third and fourth columns). 
The superscript $^{(1)}$ indicates that the fit was performed assuming 
$U(3)_l \times U(3)_e \times U(3)_d\times U(2)_u \times U(2)_q $ symmetry. 
The superscript $^{(2)}$ indicates that the constraint would 
disappear if one used the weak basis in which the down-type quark mass matrix $M_d$ is diagonal. 
See the text for details on how the bounds from ``other processes'' have been obtained.  }
  \label{tab:results}
\end{table}

Table \ref{tab:results} also shows that, due to the current deviation from the CKM unitarity, $\beta$ and meson decays prefer nonzero values for the dimension-six interactions at the $\sim 3\sigma$ level. For some of the Wilson coefficients, the size needed to address CKM unitarity is in conflict with other measurements. For example,  the current value of $\Delta_{\rm CKM}\neq 0$ could be explained with nonzero values of the four-quark Wilson coefficients, $[C_{qq}^{(1,3)}]_{1331}\simeq -0.03$. However, these relatively large coefficients are in tension with the kaon mass difference, $\Delta m_K$. Therefore, should the current anomaly in CKM unitarity be confirmed, an explanation in terms of the Wilson coefficients in Table \ref{tab:results} would require a scenario involving multiple operators at a time.

In the following subsections, we provide details
about the operator mixing, as well as the constraints on the $C_x$ from
other observables.

\subsubsection{$C^{(3)}_{q q, \ 3311} $}\label{sec:Cqq}
The most precise constraints obtained from the above analysis are for the coefficients of four-quark operators, $C^{(3)}_{q q}$. The flavor structures that contribute to $\beta$ decays contain two first-family quarks and two third-family quarks. In the context of beta decay, these quarks are up, down, top, and bottom. 
These operators primarily contribute to $C^{(3)}_{\substack{H q \\ 11}}$ via the SM Yukawa or gauge couplings. 
The corresponding one-loop contributions are illustrated in Fig.~\ref{fig:diagramy} (left and middle diagrams) and lead to the following RGE
\begin{align}
\dot C_{\substack{Hq \\ 11}}^{(3)} & \supset   2 N_c\bigl(  C^{(3)}_{\substack{qq \\ 1133}} +C^{(3)}_{\substack{qq \\ 3311}}\bigr) \left(\frac{1}{3}g_2^2-[{Y_u}^\dagger Y_u  ]_{33} \right)\ . 
\end{align}


Furthermore, these operators also generate $Q^{(3)}_{\substack{l q \\ 1111}}$ as shown in the middle 
diagram in Fig.~\ref{fig:diagramy} after attaching a lepton line (not shown) 
to the gauge boson. The RGE terms describing this contribution are
\begin{align}
\dot C_{\substack{ lq \\ 1111 }}^{(3)} & \supset \frac{2}{3} g_2^2 N_C C^{(3)}_{\substack{q q \\ 1133}} + \frac{2}{3} g_2^2 N_C C^{(3)}_{\substack{q q \\ 3311}} \ . 
 \end{align}
This, together with the bound on $\epsilon_L$, leads to
\begin{equation}
    C^{(3)}_{\substack{q q \\ 3311}} \, ,~ C^{(3)}_{\substack{q q \\ 1133}} = (4 \pm 1.3) \times 10^{-3} \, .
\end{equation}

In addition, the
$Q^{(3)}_{\substack{q q \\ 1133}}$ and $Q^{(3)}_{\substack{q q \\ 3311}}$ operators play a significant role in various other processes at tree-level, particularly in top quark production. Multiple global fits have been performed in order to constrain the operators' coefficients; see, e.g., Refs.~\cite{Buckley:2015lku, Ethier:2021bye,Kassabov:2023hbm}. 
The most stringent constraint was obtained in Ref.~\cite{ Kassabov:2023hbm}, which performed a fit assuming $U(2)_q \times U(2)_u \times U(2)_d$ symmetry.
This resulted in a constraint on the following linear combination of SMEFT coefficients, $c_{Qq}^{3,1} = C^{(3)}_{\substack{q q \\ ii33}} + \frac{1}{6}(C^{(1)}_{\substack{q q \\ i33i}}-C^{(3)}_{\substack{q q \\ i33i}})$, where the index $i$ is summed over the first and the second families. In a global analysis, the constraint on this linear combination
is $(-0.0048 \pm 0.0122)$ with $95 \% $ confidence level (CL) intervals. 
Assuming only a single Wilson coefficient from the Warsaw basis to be present at a time, one obtains
\begin{eqnarray}
    C^{(3)}_{\substack{q q \\ 3311}} \, ,~ C^{(3)}_{\substack{q q \\ 1133}}=(-0.0073 \pm 0.0053)\,,
\end{eqnarray}
at $68 \%$ CL, implying the loop level constraint from $\beta$ decays are approximately five times more sensitive.

Finally,  the  strong but basis-dependent 
constraints from  $\Delta S=1$ and $\Delta S=2$ FCNC processes are discussed in Section~\ref{sect:fcnc}.

\subsubsection{$ C^{(3)}_{lq,2233}$} \label{subsect:Clq}
The next-best constraint
arising from our analysis relates to the operator $Q_{lq}^{(3)}$, with two second-family leptons and two third-family quarks, as shown in Eq.~\eqref{eq:1}. 
This operator generates $Q^{(3)}_{\substack{H l \\ 22}}$ that characterizes the vertex for muon decay. The analogous operator involving the first generation, $Q^{(3)}_{\substack{H l \\ 11}}$, is also generated by $ C^{(3)}_{lq,1133}$
(however, as shown in Eq.~\eqref{eq:epsilonLdef} such effects cancel in $\epsilon_L^{(d,s)}$).
The  $Q^{(3)}_{\substack{l q \\ 2233}}$ contribution to $Q^{(3)}_{\substack{H l \\ 22}}$ is shown in the left diagram in Fig.~\ref{fig:diagramy}.
The  RGE term that 
generates the dominant contribution reads:
\begin{align}
\dot C_{\substack{Hl  \\ 22}}^{(3)} & \supset 2N_c  C^{(3)}_{\substack{lq \\ 2233 }} \left(\frac{1}{3}g_2^2-  [Y_u^\dagger Y_u]_{33}\right)\,.
\end{align}
After solving the above RGE and imposing the CC constraints, we obtain
\begin{equation}
    C^{(3)}_{\substack{l q \\ 2233}} = (-8 \pm 2.5) \times 10^{-3} \, .
\end{equation}

The same operator also plays a role in B decays at a tree level. In a single-coupling analysis, the comparison of $b\to c \mu \nu$ and $b\to c e \nu$ transitions can constrain the relevant coefficient, leading to an allowed range of $[-0.0005,0.022]$
at $\mu = 1$ TeV \cite{Jung:2018lfu}. This bound is approximately one order of magnitude weaker compared to what was obtained from $\beta$ decay. 

A stronger constraint can be obtained from $B\to K\ell\ell$ decays, 
to which $C^{(3)}_{\substack{l q \\ 2233}}$ contributes at tree level after performing the 
CKM rotation to the mass basis in $Q^{(3)}_{\substack{l q \\ 2233}}$.
Ref.~\cite{Garosi:2023yxg} performed a fit at $\mu=1$ TeV, assuming that only operators involving top quarks are present at $\mu=1$ TeV, while the lepton flavor structure was left arbitrary. The dominant constraint comes from the Lepton Flavor Universality ratio
\begin{equation}
    R_K = \frac{{\rm Br}(B^{+} \xrightarrow{} K^{+} \mu^{+} \mu^{-})}{{\rm Br}(B^{+} \xrightarrow{} K^{+} e^{+} e^{-})} \ .
\end{equation}
and implies  $C^{(3)}_{\substack{l q \\ 2233}} =  (-2.4 \pm 2.1)\times 10^{-4}$
for the scenario when only one operator is present at a time.
This bound has a smaller error and mean value compared to the constraint from low-energy CC observables.

\subsubsection{$ C_{ll,1122}$}
These Wilson coefficients generate $C_{\substack{l l \\ 1221}}$
through loop diagrams that involve the gauge coupling, as shown in the right panel of Fig.~\ref{fig:diagramy}. This is characterized by the RGEs
\begin{align}
\dot C_{\substack{ll  \\ 1221}} \supset  6 \ C_{\substack{ll  \\ 1122}} g_2^2\ , ~~~~~~~~ \dot C_{\substack{ll  \\ 2112}} \supset  6 \ C_{\substack{ll  \\ 2211}} g_2^2 \ , 
\end{align}
leading to the constraint
\begin{equation}
    C_{\substack{l l \\ 1122}} = C_{\substack{l l \\ 2211}} = (-1.8 \pm 0.5) \times 10^{-2} \ .
\end{equation}

The above-mentioned operators also induce muon pair production at tree level. 
Ref.~\cite{Falkowski:2015krw} performed a global fit that incorporated experimental data from various sources, including Z boson measurements at LEP-1, W boson mass and decay measurements, muon and tau decays, and lepton pair production at LEP-2. Under the assumption that only one operator is present at a time, this led to an allowed range of $(-1.8 \pm 2.9) \times 10^{-3} $ at $1$ TeV with $68\%$ CL, 
comparable to the bound from low-energy CC processes.

\subsubsection{$ C^{(3)}_{qq,1331}$}
The operator  $Q^{(3)}_{\substack{q q}}$, discussed in Section \ref{sec:Cqq}, reappears in the analysis, albeit with different particle content. 
In particular, $C^{(3)}_{qq,1331}$ contributes to both $C^{(3)}_{\substack{H q \\ 11}}$ and $C^{(3)}_{\substack{l q \\ 1111}}$ 
via the left and middle diagrams of  Fig.~\ref{fig:diagramy}, 
\begin{align}
\dot C_{\substack{Hq \\ 11}}^{(3)} & \supset  \bigl( C^{(3)}_{\substack{qq \\ 1331}} + C^{(3)}_{\substack{qq \\ 3113}} \bigr) \left([{Y_u}^\dagger Y_u]_{33}-\frac{1}{3}g_2^2\right) \,,\\
\dot C_{\substack{ lq \\ 1111 }}^{(3)} &  \supset - \frac{1}{3} g_2^2 C^{(3)}_{\substack{q q \\ 1331}} - \frac{1}{3} g_2^2 C^{(3)}_{\substack{q q \\ 3113}} \ . 
 \end{align}
These RGEs lead to the following limits from $\beta$ decays,
\begin{equation}
    C^{(3)}_{\substack{q q \\ 1331}} \, ,~ C^{(3)}_{\substack{q q \\ 3113}} = (-3.0 \pm 0.9) \times 10^{-2} \ .
\end{equation}

The above effective couplings modify  top quark production processes.
Ref.~\cite{ Kassabov:2023hbm} 
obtained the following constraints on several linear combinations of Wilson coefficients,
\begin{subequations}
\label{4fermi-operators}
\begin{eqnarray*}
& &\text{Linear combination} ~~~~~~~ \qquad \qquad \quad  \text{Constraint} \\
& &c_{Qq}^{1,8} = C^{(1)}_{\substack{q q \\ i33i}} + 3C^{(3)}_{\substack{q q \\ i33i}}, ~~~~~~~~~~~~~~~~~~~~~~ 0.121 \pm 0.30 \ , \\
& &c_{Qq}^{1,1} = C^{(1)}_{\substack{q q \\ ii33}} + \frac{1}{6}C^{(1)}_{\substack{q q \\ i33i}}+\frac{1}{2}C^{(3)}_{\substack{q q \\ i33i}} ,~~~~~~~~~~0.30 \pm 0.61  \ ,\\
& & c_{Qq}^{3,8} = C^{(1)}_{\substack{q q \\ i33i}} - C^{(3)}_{\substack{q q \\ i33i}} ,~~~~~~~~~~~~~~~~~~~~~~-0.12 \pm 0.30 \ ,
\end{eqnarray*}
\end{subequations}
as well as the previously mentioned constraint on $c_{Qq}^{3,1}$, see Section \ref{sec:Cqq}. Evaluating these results, assuming that only one Wilson Coefficient from the Warsaw basis is present at a time,  we extract  
$ C^{(3)}_{\substack{q q \\ 1331}} \approx (-0.042 \pm 0.024)$
at $68 \%$ CL, which is 
a factor of a few weaker than the result from $\beta$ decays. 
Finally,  our analysis of the 
constraints from  $\Delta S=1$ and $\Delta S=2$ FCNC processes is discussed in Section~\ref{sect:fcnc}.

\subsubsection{$ C^{(1)}_{qq,1331}$}
The final operator coefficient from Table~\ref{tab:results} we discuss here is similar to that of the previous section but with a different isospin dependence.
 It contributes primarily to the operators $Q^{(3)}_{\substack{Hq \\ 11}}$ through the loop diagrams involving a Yukawa interaction in Fig.\ \ref{fig:diagramy}. The RG equation 
\begin{equation}
\dot C^{(3)}_{\substack{Hq \\ 11}} \supset \left(C^{(1)}_{\substack{q q \\ 1331}}+C^{(1)}_{\substack{q q \\ 3113}}\right) \left(\frac{1}{3} g_2^2-[{Y_u}^\dagger Y_u  ]_{33} \right)\ 
\end{equation}
leads to the following limit from low-energy CC processes: 
\begin{equation}
    C^{(1)}_{\substack{q q \\ 1331}} \, ,~ C^{(1)}_{\substack{q q \\ 3113}} = (3.1 \pm 0.9) \times 10^{-2} \ .
\end{equation}

Direct constraints on this coefficient have been obtained in Ref.~\cite{ Kassabov:2023hbm} 
through top quark production. 
Evaluating the fit while only one Wilson coefficient of the Warsaw basis was present at a time, at $68 \%$ CL, we extract
$C^{(1)}_{\substack{q q \\ 1331}}  \approx  -0.035 \pm 0.027$. This result is a factor of a few weaker than the limit obtained from $\beta$ decay, albeit with a similar central value.
We discuss the 
constraints from  $\Delta S=1$ and $\Delta S=2$ FCNC processes in Section~\ref{sect:fcnc}.

\subsubsection{FCNC processes}
\label{sect:fcnc}
As mentioned above, the limits on the Wilson coefficient that arise from beta decay at the loop level are competitive with those obtained from other processes that are induced at the tree level. 
Here, we discuss several other probes of the Wilson coefficients in more detail, with particular emphasis on Flavor-Changing Neutral Current (FCNC) processes. FCNCs  are suppressed in the Standard Model due to the GIM mechanism~\cite{Glashow:1970gm}, which is generally violated in 
 New Physics models and within SMEFT. These observables are therefore highly sensitive to contributions from physics beyond the Standard Model, given the low SM background and high experimental precision. 

As discussed in section \ref{subsect:Clq},
probes of $\Delta F=1$ transitions are able to constrain a number of SMEFT operators at the tree level.
In this section, we instead focus on the constraints from $\Delta F=1$ FCNC processes that are induced at the loop level and from $\Delta F=2$ observables.
To investigate the former, we look for the FCNC dimension-six coefficients ($C_{FCNC}$) that could be generated by the same $C_x$ coefficients that generate $C_{\beta}$, allowing us to compare the resulting limits. The relevant RGEs can be written schematically as
\begin{subequations}
\begin{eqnarray}
    \dot{C}_{\beta} &=& \gamma_{\beta x} \ C_{x} \ , \\
    \dot{C}_{FCNC} &=& \gamma_{FCNC x} \ C_{x} \ .
\end{eqnarray}
\end{subequations}
We focus on processes that can be calculated with negligible theoretical uncertainty, specifically, the rare decays of kaons and $B$ mesons, taking into account the effects of the rotation to the mass basis. 
The strongest constraints arise from the rare $\Delta S=1$ FCNC process $K \to \pi \nu \bar \nu$, which is controlled by the LEFT coefficient 
$[L_{VLL}^{\nu d}]_{\ell \ell 12}$,  currently bound to be smaller than  
$0.97 \times  10^{-4} \,  {\rm TeV}^{-2}$ \cite{Buras:2015qea,Allwicher:2023shc}.
In terms of the SMEFT Wilson coefficients in our basis,  we have 
\begin{eqnarray}
   v^2 [L_{VLL}^{\nu d}]_{\ell \ell 12} &=& \tilde C_{\ell \ell 12} + \lambda \left( \tilde C_{\ell \ell 11} - \tilde C_{\ell \ell 22} \right) + O(\lambda^2) 
    \nonumber \\
    \tilde C_{\ell \ell ij} &=&  C_{\substack{lq \\ \ell \ell ij}}^{(1)} - C_{\substack{lq\\ \ell \ell ij}}^{(3)} + \delta_{\ell \ell} C_{\substack{Hq\\ij}}^{(1)} + \delta_{\ell \ell} 
    C_{\substack{Hq\\ij}}^{(3)}~.  
     \end{eqnarray}
For the running of $C_{qq}^{(1),(3)}$ into these couplings we find 
\begin{eqnarray}
    \tilde C_{\ell \ell 11}&\simeq & \left( -0.099 \ C_{\substack{qq\\1133}}^{(1)}-0.0034\ C_{\substack{qq\\1331}}^{(1)}+0.098 
    \ C_{\substack{qq\\1133}}^{(3)}-0.054\ C_{\substack{qq\\1331}}^{(3)} \right) \delta_{\ell \ell} \ ,
\end{eqnarray}
while $\tilde C_{12,22}$ are much smaller than $\tilde C_{11}$. 
The limit on $[L_{VLL}^{\nu d}]_{\nu_\ell \nu_\ell 12}$ implies 
\begin{eqnarray}
    C_{\substack{qq\\1133}}^{(1)}<2.0 \cdot 10^{-5}\,,\qquad C_{\substack{qq\\1331}}^{(1)}< 0.57\cdot 10^{-3}\,,\\
        C_{\substack{qq\\1133}}^{(3)}<2.0 \cdot 10^{-5}\,,\qquad C_{\substack{qq\\1331}}^{(3)}< 0.37 \cdot 10^{-4}\, . 
\end{eqnarray}

Apart from measurements of rare $\Delta F=1$ processes that violate flavor by one unit, stringent constraints arise from $\Delta F=2$ observables, such as the $B_{d,s}-\bar B_{d,s}$ and $K_L-K_S$ mass differences. These mass differences are suppressed due to the GIM mechanism in the SM, which implies that the contributions from the four-quark operators $Q_{qq}^{(1),(3)}$ can be significant as they arise at the tree level, after rotating to the mass basis.
To evaluate the constraints from $\Delta m_{B_{d,s}}$, we compare the sum of the SM \cite{Buras:2022wpw} and BSM contributions \cite{Aebischer:2020dsw} with the experimental determinations \cite{ParticleDataGroup:2022pth}. The SM prediction of $\Delta m_K$ is subject to larger theoretical uncertainties. Although the short-distance contributions are known to about $30\%$, the long-distance effects are poorly known \cite{Brod:2011ty,Buras:2018wmb}. In fact, preliminary lattice-QCD results are currently subject to significant systematic  uncertainties~\cite{Wang:2022lfq}. Therefore, to set conservative constraints, we demand that the BSM contributions cannot be larger than the experimental determination, $\Delta m_K^{\rm BSM}\leq \Delta m_K^{\rm exp}$. 
Although it comes with significant uncertainties, the kaon mass difference is still the most sensitive probe of the $C_{qq}^{(1),(3)}$ coefficients, while the limits from $\Delta m_{B_d}$ are a factor of a few weaker. 
Quantitatively, the $\Delta S=2$ limits imply $|C^{(3)}_{\substack{q q \\ 1 1 3 3}}|, |C^{(1)}_{\substack{q q \\ 1 3 3 1}}|, |C^{(3)}_{\substack{q q \\ 1 3 3 1}}|  < 0.0009$.
These constraints are weaker than the ones from $K \to \pi \nu \bar \nu$, 
except for $C_{\substack{qq\\1331}}^{(1)}$.

At face value, the 
$\Delta S=1$ and 
$\Delta S=2$ observables then lead to some of the strongest constraints on the $C_{qq}^{(1,3)}$ coefficients in Table \ref{tab:results}. It should be noted, however, that these flavor-violating probes strongly depend on the assumed flavor structure of the Wilson coefficients. For example, defining the operators in the basis where $M_d$ is diagonal, instead of using Eq.\ \eqref{eq:basis}, would lead to severely weakened 
$\Delta F=1$ and 
$\Delta F=2$ constraints \cite{Aebischer:2020dsw}, while those from low-energy CC processes and $pp\to t\bar t$ would be mostly unaffected.

\section{Simplified models above the $\Lambda$ scale}\label{sec:models}
Here we briefly discuss the kind of BSM scenarios that could give rise to the operators discussed in the previous section. 
Above the $\Lambda$ scale, such simplified scenarios can be constructed by introducing new heavy fields with all possible spin- and gauge representations.
The complete list of new particles
and their tree-level contributions to the SMEFT operators up to dimension-six are provided in Ref.~\cite{deBlas:2017xtg}. This matching procedure allows us to express the SMEFT Wilson coefficients as functions of the couplings and masses
in the appropriate simplified model. With the aid of the provided dictionary, we can readily identify all heavy fields that can generate the Wilson coefficients for which we established constraints. The results are summarized in Table \ref{tab:heavy_fields}. 

There are two heavy fields, marked with an asterisk in Table \ref{tab:heavy_fields}, which exclusively contribute to the $Q^{(3)}_{\substack{q q}}$ and $Q^{(1)}_{\substack{q q}}$ operators that were constrained from $\beta$ decay at the loop level. These heavy fields are a scalar boson $\Upsilon$, which is a color sextet and a weak isotriplet, along with a vector boson $\mathcal{H}$, which is a color octet and a weak isotriplet. 
The scalar field can play a role in explanations of baryogenesis \cite{Arnold:2012sd} and contributes to processes at the LHC \cite{Zhan:2013sza}. The vector field $\mathcal{H}$ can appear as the gauge boson of an enlarged gauge group, $SU(6)$, which is broken down to the SM  \cite{delAguila:2010mx}, or as a resonance arising from strongly coupled BSM physics \cite{Krause:2018cwe}.
Their contributions to the $C_x$ operator coefficients, assuming only one multiplet at a time, read
\begin{subequations}
\label{eq:ModelMatching}
\begin{eqnarray}
&& 
C^{(1)}_{\substack{q q \\ i j k l}} = \frac{3(y_{\Upsilon})_{lj} (y_{\Upsilon})^*_{ki}}{4M_{\Upsilon}^2} \ , ~~~~~~~~~~~~~~~
C^{(3)}_{\substack{q q \\ i j k l}} = \frac{3(y_{\Upsilon})^*_{ki} (y_{\Upsilon})_{jl}}{M_{\Upsilon}^2} \ , \\
&& 
C^{(1)}_{\substack{q q \\ i j k l}} = -\frac{3(g_{\mathcal{H}})_{kj}(g_{\mathcal{H}})_{il}}{32M^2_{\mathcal{H}}} \ , ~~~~~~~~~~~~
C^{(3)}_{\substack{q q \\ i j k l}} = \frac{(g_{\mathcal{H}})_{kl}(g_{\mathcal{H}})_{ij}}{48M^2_{\mathcal{H}}} + \frac{(g_{\mathcal{H}})_{kj}(g_{\mathcal{H}})_{il}}{32M^2_{\mathcal{H}}}  \ . 
\end{eqnarray}
\end{subequations}
where $M_{\Upsilon}$ and $M_{\mathcal{H}}$ are the masses of the heavy fields and $y_{\Upsilon}$ and $y_{\mathcal{H}}$ are their couplings to left-handed quarks. With respect to the flavor structure of these coefficients, it becomes apparent that we require significant couplings of $(y_{\Upsilon})_{13}$, $(y_{\Upsilon})_{31}$, $(y_{\mathcal{H}})_{11}$, $(y_{\mathcal{H}})_{33}$, $(y_{\mathcal{H}})_{13}$, $(y_{\mathcal{H}})_{31}$  to address the Cabibbo angle anomaly. For sufficiently small masses of heavy fields, these couplings can be examined in various
processes at a tree level, particularly in quark top production.
In addition, the needed $\Upsilon$ couplings involve different quark generations, implying significant contributions to FCNC processes in case the diagonal couplings are nonzero as well, $(y_{\Upsilon})_{ii}\neq0$.

 \begin{table}[!h]
  \begin{adjustbox}{center}
  \setlength\extrarowheight{-1pt}
  \def\arraystretch{1}  
    \begin{tabular}{cccccccc}
      \toprule
      $Q_{x}$ operator & \multicolumn{6}{c}{Heavy field}  \\
      \midrule
      $Q^{(3)}_{\substack{q q}}$  & $ \omega_1$ & $\zeta$  &   $\Omega_1$ & $\Upsilon^*$ & $\mathcal{W}$  & $\mathcal{G}$, & $\mathcal{H^*}$    \\
     Irrep & $(3,1)_{-\frac{1}{3}}$ &$ (3,3)_{-\frac{1}{3}}$ & $(6,1)_{\frac{1}{3}}$&$(6,3)_{\frac{1}{3}}$ & $(1,1)_0$ & $(8,1)_0 $ & $(8,3)_0 $ \\ \midrule
      $Q^{(1)}_{\substack{q q}}$  & $ \omega_1$ & $\zeta$  &   $\Omega_1$ & $\Upsilon^*$ & $\mathcal{B}$  & $\mathcal{G}$, & $\mathcal{H^*}$    \\
      Irrep & $(3,1)_{-\frac{1}{3}}$ &$ (3,3)_{-\frac{1}{3}}$ & $(6,1)_{\frac{1}{3}}$&$(6,3)_{\frac{1}{3}}$ & $(1,3)_0$ & $(8,1)_0 $ & $(8,3)_0 $\\\midrule
      $Q^{(3)}_{\substack{l q}}$  &  $ \omega_1$ & $\zeta$  & $\mathcal{W}$ &  $\mathcal{U}_2$  &    $\mathcal{X}$      \\
       Irrep & $(3,1)_{-\frac{1}{3}}$ &$ (3,3)_{-\frac{1}{3}}$ & $(1,3)_0$ & $(3,1)_0$ & $(3,3)_{\frac{2}{3}}$ \\ \midrule
      $Q_{\substack{l l}}$   & $\mathcal{S}_1$ &  $\Xi$  & $\beta$ & $\mathcal{W}$ \\
      Irrep & $(1,1)_1$ & $(1,3)_1$ & $(1,1)_0$ & $(1,3)_0$  \\

      \bottomrule
    \end{tabular}
  \end{adjustbox}
  \caption{Single-field extensions of the Standard Model generating $Q_x$ operators at tree level. Here we follow the 
  notation of Ref.~\cite{deBlas:2017xtg}. The asterisks indicate fields that only generate the operators that contribute to the $\beta$ decays at the loop level.}
  \label{tab:heavy_fields}
\end{table}

\section{Conclusions}\label{sec:conclusions}

We performed a loop-level analysis of 
low-energy (semi)leptonic charged current processes 
in the Standard Model Effective Field Theory truncated at dimension six.
Using the one-loop anomalous dimensions and the corresponding renormalization group equations  we have expressed the  low-scale effective couplings  
affecting $\beta$ decays,  $\epsilon_\alpha (\mu_W)$, in terms of the SMEFT Wilson coefficients at the new-physics scale, $\Lambda$.
We find that the RGE running generates new significant contributions only for the left-handed effective coupling $\epsilon_L$ that shifts the Standard Model $(V-A)\times (V-A)$ operator, 
while providing only small shifts to the tree-level matching conditions (see Eqs.~\eqref{eq:tree_level_matching}) for the right-handed current, scalar,   and tensor operators. 

Our main results are given by Eqs.~\eqref{eq:epsilonL} and \eqref{eq:epsilonLs}, 
which display $\epsilon_L^{(d,s)}(\mu_W)$ in terms of the SMEFT Wilson coefficients $C_n (\Lambda)$, 
keeping only the effective couplings that contribute with coefficients larger than $10^{-3}$.
Eq.~\eqref{eq:epsilonL}, coupled with the experimental input on $\epsilon_L^{(d)} (\mu_W)$ from a
recent global analysis of beta decays and other precision measurements~\cite{Cirigliano:2023nol}, 
provides the general constraint from low-energy charged-current processes on SMEFT effective couplings, at the leading logarithmic accuracy. 
Turning on the SMEFT couplings one at a time at the reference scale $\Lambda=5$ TeV, we obtain the constraints 
displayed in Fig.~\ref{fig:wyniki} and Table~\ref{tab:results}, which we compare with the sensitivity of other observables. 
We find that low-energy charged-current processes are generally competitive and
provide bounds that do not depend on assumptions about flavor symmetries imposed on the SMEFT couplings. 
We find that the Wilson coefficients of operators involving the third generation, such as $Q_{qq}^{(1),(3)}$ and $Q_{lq}^{(3)}$, are stringently constrained by low-energy CC processes to be at  
the permille level, corresponding to a reach of $\Lambda\simeq 8$ TeV.
While the current tension with first-row CKM unitarity points to a nonzero combination of SMEFT couplings, 
further theoretical and experimental scrutiny may shift the central value and will certainly improve confidence
in the uncertainties, and thus the robustness of the resulting constraints.

Finally, we have also identified single-field extensions of the Standard Model that  
contribute to the SMEFT operators, which generate significant effects in beta decays via loops (see Table~\ref{tab:heavy_fields}).  
It is interesting to note that the scalar field $\Upsilon$ and vector field $\cal H$ at tree level only generate the $Q_{qq}^{(1),(3)}$ operators. 
These extensions therefore provide simplified UV scenarios in which the first contributions to low-energy charged-current processes appear at loop level.
\\

{\bf Acknowledgments -- }  
We thank Emanuele Mereghetti and Tom Tong for the helpful discussions and comments on the manuscript. 
This work is supported by the U.S. DOE Office of Nuclear Physics under Grant No. DE-FG02-00ER41132  
and by the DOE Topical Collaboration ``Nuclear Theory for New Physics", award No. DE-SC0023663. 

\newpage
\appendix
\section{Numerical results for the RG evolution} \label{ap:results}
{\setlength{\tabcolsep}{5.25pt}
\footnotesize
\renewcommand*{\arraystretch}{1.8}
\begin{longtable}{@{}l*{10}{S[table-format=5.2,table-space-text-pre=(, table-space-text-post=)]}S[table-format=3.2,table-space-text-pre=(, table-space-text-post=)]@{}}
    \caption{Results for $U_{ik}(\mu_W, \Lambda)$, which describes the ratio of $C_{\beta i} (\mu_W)$ at low electroweak scale $\mu_W = 0.246$~TeV to $C_{xk} (\Lambda)$ at high scale $\Lambda=5$~TeV as defined in \eqref{eq:CbetaRG}.
    In the table below, the rows correspond to the index $k$ while the columns correspond to the index $i$.
    We report here only the entries with $ |U_{ik}(\mu_W, \Lambda) |> 10^{-3}$.
    }
    \label{tab:U_results}\\
    \toprule
     $C_x$     & $C^{(3)}_{\substack{H l \\ 1 1}}$ & $C^{(3)}_{\substack{H l \\ 2 2}}$ & $C^{(3)}_{\substack{H q \\1 1}}$ & $C^{(3)}_{\substack{l l \\ 1 2 2 1}}$ & $C^{(3)}_{\substack{l l \\ 2 1 1 2 }}$ & $C^{(3)}_{\substack{l q \\1 1 1 1}}$  \\
      \midrule
      $C^{(3)}_{\substack{q q \\1 1 1 1}}$ & 0        & 0        & -0.0261   & 0        & 0        & -0.0262  \\
      $C^{(3)}_{\substack{q q \\1 1 2 2}}$ & 0        & 0        & -0.0153  & 0        & 0        & -0.0154  \\
      $C^{(3)}_{\substack{q q \\1 1 3 3}}$ & -0.0010  & -0.0010  & 0.0819    & 0        & 0        & -0.0154  \\
      $C^{(3)}_{\substack{q q \\1 2 2 1}}$ & 0        & 0        & 0.0025   & 0        & 0        & 0.0025  \\
      $C^{(3)}_{\substack{q q \\1 3 3 1}}$ & 0        & 0        & -0.0113   & 0        & 0        & 0.0025  \\
      $C^{(3)}_{\substack{q q \\2 1 1 2}}$ & 0        & 0        & 0.0025   & 0        & 0        & 0.0025  \\
      $C^{(3)}_{\substack{q q \\2 2 1 1}}$ & 0        & 0        & -0.0153   & 0        & 0        & -0.0154  \\
      $C^{(3)}_{\substack{q q \\2 2 3 3}}$ & -0.0010  & -0.0010  & -0.0010  & 0        & 0        & 0       \\
      $C^{(3)}_{\substack{q q \\3 1 1 3}}$ & 0        & 0        & -0.0113   & 0        & 0        & 0.0025  \\
      $C^{(3)}_{\substack{q q \\3 3 1 1}}$ & -0.0010  &  -0.0010 & 0.0818    & 0        & 0        & -0.016  \\
      $C^{(3)}_{\substack{q q \\3 3 2 2}}$ & -0.0010  &  -0.0010 & -0.0010  & 0        & 0        & 0       \\
      $C^{(3)}_{\substack{q q \\3 3 3 3}}$ & -0.0021  &  -0.0021 & -0.0021  & 0        & 0        & 0       \\
      $C^{(1)}_{\substack{q q \\1 1 1 1}}$ & -0.0158        & 0        & -0.0037  & 0        & 0        & -0.0037 \\
      $C^{(1)}_{\substack{q q \\1 1 3 3}}$ & 0        & 0        & -0.0020   & 0        & 0        & 0       \\
      $C^{(1)}_{\substack{q q \\1 2 2 1}}$ & 0        & 0        & -0.0020  & 0        & 0        & -0.0020 \\
      $C^{(1)}_{\substack{q q \\1 3 3 1}}$ & 0        & 0        & 0.0110    & 0        & 0        & -0.0020 \\
      $C^{(1)}_{\substack{q q \\2 1 1 2}}$ & 0        & 0        & -0.0022  & 0        & 0        & -0.0021 \\
      $C^{(1)}_{\substack{q q \\3 1 1 3}}$ & 0        & 0        & 0.012    & 0        & 0        & -0.0021 \\
      $C^{(1)}_{\substack{q q \\3 3 1 1}}$ & 0        & 0        & -0.001   & 0        & 0        & 0       \\
      $C^{(3)}_{\substack{l q \\1 1 1 1}}$ & -0.0158   & 0        & -0.0051        & -0.0156   & -0.0156   & 1.0311  \\
      $C^{(3)}_{\substack{l q \\1 1 2 2}}$ & -0.0158   & 0        & 0        & -0.0160   & -0.0160   & -0.0157  \\
      $C^{(3)}_{\substack{l q \\1 1 3 3}}$ & 0.0834    & 0        & 0        & -0.0156   & -0.0156   & -0.0157  \\
      $C^{(3)}_{\substack{l q \\2 2 1 1}}$ & 0        & -0.0158   & -0.0051  & -0.0156   & -0.0156   & -0.0051 \\
      $C^{(3)}_{\substack{l q \\2 2 2 2}}$ & 0        & -0.0158   & 0        & -0.0156   & -0.0156   & 0       \\
      $C^{(3)}_{\substack{l q \\2 2 3 3}}$ & 0        & 0.0834    & 0        & -0.0156   & -0.0156  & 0       \\
      $C^{(3)}_{\substack{l q \\3 3 1 1}}$ & 0        & 0        & -0.0056  & 0        & 0        & -0.0051 \\
      $C_{\substack{l l \\1 1 1 1}}$       & -0.0051  & 0        & 0        & -0.0051  & -0.0051  & -0.0052 \\
      $C_{\substack{l l \\1 1 2 2}}$       & 0        & 0        & 0        & -0.049   & 0        & 0       \\
        $C_{\substack{l l \\1 2 2 1}}$       & -0.0026        & -0.0026         & 0        & 1.007        & -0.0052   & -0.0026        \\
      $C_{\substack{l l \\1 3 3 1}}$       & -0.0026  & 0        & 0        & -0.0027  & -0.0026  & -0.0027 \\
       $C_{\substack{l l \\2 1 1 2}}$       & -0.0026        & -0.0026         & 0        & -0.0052        & 1.007   & -0.0026        \\
      $C_{\substack{l l \\2 2 1 1}}$       & 0        & 0        & 0        & 0        & -0.049   & 0       \\
      $C_{\substack{l l \\2 2 2 2}}$       & 0        & -0.0054  & 0        & -0.0051  & -0.0051  & 0       \\
      $C_{\substack{l l \\2 3 3 2}}$       & 0        & -0.0027  & 0        & -0.0026  & -0.0026  & 0       \\
      $C_{\substack{l l \\3 1 1 3}}$       & -0.0026  & 0        & 0        & -0.0026  & -0.0026  & -0.0027 \\
      $C_{\substack{l l \\3 2 2 3}}$       & 0        & -0.0027  & 0        & -0.0026  & -0.0026  & 0       \\
      $C^{(3)}_{\substack{H q \\1 1}}$     & -0.0143   & -0.0143   & 0.9257   & 0        & 0        & -0.0026       \\
      $C^{(3)}_{\substack{H q \\2 2}}$     & -0.0143  & -0.0143   & -0.0141   & 0        & 0        & 0       \\
      $C^{(3)}_{\substack{H q \\3 3}}$     & -0.0142   & -0.0142   & -0.0140   & 0        & 0        & 0       \\
      $C^{(3)}_{\substack{H l \\1 1}}$     & 0.957  & -0.005   & -0.005  & -0.003        & -0.003        & -0.003       \\
      $C^{(3)}_{\substack{H l \\2 2}}$     & -0.0053  & 0.957   & -0.0053  & -0.003        & -0.003        & 0       \\
      $C^{(3)}_{\substack{H l \\3 3}}$     & -0.0048  & -0.0048   & -0.0048  & 0        & 0        & 0       \\
      $C_{H \Box}$                         & -0.0013  & -0.0013  & -0.0013  & 0        & 0        & 0       \\
\end{longtable}%
 }
\newpage
\bibliographystyle{JHEP}
\bibliography{references,references2}

\end{document}